\documentclass[a4paper]{mem}
\usepackage{natbib}
\usepackage{graphicx}
\usepackage[a4paper]{hyperref}
\idline{}{1}
\newcommand{\etal}{{et~al.}}

\begin{document}

\title{The {\sc Planck} Low Frequency Instrument}

\author{N.~Mandolesi\inst{1}, C.~Burigana\inst{1}, R.~C.~Butler\inst{1},
F.~Cuttaia\inst{1}, A.~De~Rosa\inst{1}, F.~Finelli\inst{1}, 
E.~Franceschi\inst{1}, A.~Gruppuso\inst{1}, M.~Malaspina\inst{1}, 
G.~Morgante\inst{1}, G.~Morigi\inst{1}, 
L.~Popa\inst{1}, 
M.~Sandri\inst{1}, 
L.~Stringhetti\inst{1},
L.~Terenzi\inst{1},~~L.~Valenziano\inst{1},~~\and~~F.~Villa\inst{1}
}
\institute{$^1$CNR-INAF/IASF, Sezione di Bologna,
           Via Gobetti, 101, I-40129 Bologna, Italy \\ 
\email{\textit{surname}@bo.iasf.cnr.it} 
}

\abstract{
{\sc Planck} is the third generation of mm-wave instruments designed for space observations of the cosmic microwave background (CMB) anisotropies within the new Cosmic Vision 2020 ESA Science Program. 
{\sc Planck} will map the whole sky with unprecedented sensitivity, angular resolution, and frequency co\-verage, and it likely leads us to the final comprehension of the CMB anisotropies. 
The Low Frequency Instrument (LFI), operating in the 30 $\div$ 70 GHz range, 
is one of the two instruments onboard {\sc Planck} satellite, 
sharing the focal region of a 1.5 meter off-axis dual reflector telescope together with the High Frequency Instrument (HFI) operating at 100 $\div$ 857 GHz. We present LFI and discuss the major instrumental systematic effects that could degrade the measurements and the solutions adopted 
in the design and data analysis phase in order to adequately reduce and 
control them. 
{\it (On behalf of LFI Consortium)}
\keywords{Space mission, experimental cosmology, cosmic microwave background}
}

\authorrunning{N.~Mandolesi et al.}
\titlerunning{The {\sc Planck} Low Frequency Instrument}
\maketitle

\section{Introduction}
{\sc Planck} represents the third generation of mm-wave instruments 
designed for space observations of the cosmic microwave background (CMB) 
anisotropies within the new Cosmic Vision 2020 ESA Science Program.
Following the present NASA's mission Wilkinson Microwave Anisotropy 
Probe (Bennett et al. 2003)
{\sc Planck} will map the whole sky with unprecedented sensitivity (the average sensitivity on a pixel of {\sc fwhm} side in the measurement of the temperature anisotropy is about two parts per million), angular resolution ({\sc fwhm} from about 30 down to 5 arcmin), and frequency coverage, and it likely leads us to the final comprehension of the CMB anisotropies 
(J. Tauber 2004, {\it this Meeting}). 
The Low Frequency Instrument (LFI, Mandolesi et al. 1998), operating in the 30 $\div$ 70 GHz range, is one of the two instruments onboard {\sc Planck} satellite, sharing the focal region of a 1.5 meter off-axis dual reflector telescope together with the High Frequency Instrument (HFI, Puget et al. 1998) o\-perating at 100 $\div$ 857 GHz. 
LFI consists of four main units: the front end unit 
(FEU, Sect.~2), 
the back end unit (BEU, 
Sect.~3), 
the radiometer electronics box assembly 
(REBA, 
Sect.~4), 
and the sorption cooler system 
(SCS, Sect.~5).
We describe how these units have been conceived, 
as well as the major instrumental systematic effects that could 
degrade the measurements and the solutions adopted in the 
design phase in order to adequately reduce and control them. 
In addition, the removal of systematic effects and the LFI data processing 
center (DPC) are mentioned and described in 
Sects.~6 and~7.

\begin{figure}[h]
\centering
{\includegraphics[width=6.5cm]{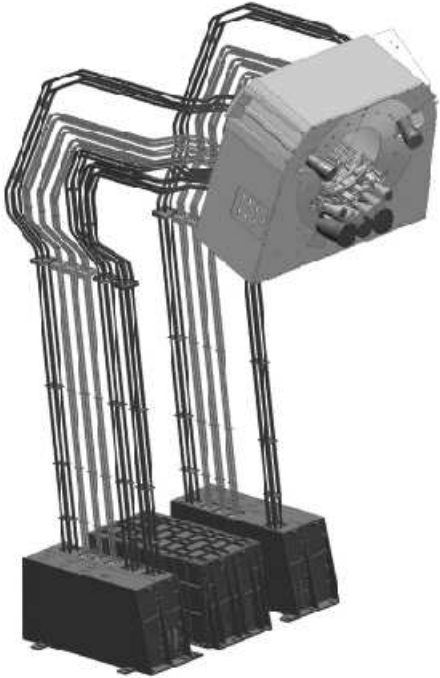}}
\caption{The Low Frequency Instrument: the front end unit is at the top, whereas the back end unit, the radiometer electronics box assembly, and sorption cooler system are in the lower part.}
\label{lfis}
\end{figure}

\section{The Front End Unit}
\label{feu}

\subsection{Feed horns and OMTs}
LFI is coupled to the {\sc Planck} telescope by an array of conical dual profiled 
corrugated feed horns 
(Villa et al. 2002).
Dual profiled corrugated horns have been selected as the best design in terms of shape of the main lobe, very low level of cross polarization and sidelobe, control of the phase centre location, low weight and compactness. 
In addition, the electromagnetic field inside the horn propagates with low attenuation and low return loss. 
The ortho mode transducers (OMTs) separate the orthogonal polarizations with minimal losses and cross-talk. 
The straylight, or the unwanted off-axis radiation contributing to the observed signal, represents one of the major sistematic effects in {\sc Planck}/LFI and can be controlled optimizing the feed horn design, since from the latter depends the illumination of the mirrors. A trade-off between angular resolution (the more the primary mirror is illuminated, the best is the angular resolution) and straylight rejection (the more the mirrors are illuminated, the worst is the straylight rejection) has been obtained for each feed horn coupled with the {\sc Planck} telescope. This optimization has been performed computing the full optical response of several realistic feed horn patterns 
(Sandri et al. 2003)
and convolving the full pattern with the sky signal by considering the observational stra\-tegy 
(Burigana et al. 2003)
in order to calculate the straylight contamination for each feed model analized. As a result, the LFI feed horns have different designs (i.e. inner corrugation profile), depending from their location on the focal surface, and the corresponding angular resolution achieved is the best one, satisfying the straylight rejection requirement.
\subsection{Hybrids, phase swithes and amplifiers}
The OMTs are followed by blocks conta\-i\-ning hybrid couplers and amplifiers (inclu\-ding phase switches and output hybrids), all cooled to 20K by the H$_2$ sorption cooler system. 
This front-end is designed to minimize the $1/f$ noise in the radiometer 
(one of the most important potential source of systematic effects) while 
maintaining low thermal noise 
(Seiffert et al. 2002, Mennella et al. 2003). Each block contains two 
hybrid 
couplers: each hybrid has two inputs, one of which sees the sky, the other one looks at the 4K reference load through a small rectangular horn. The hybrid coupler combines the signals from the sky and cold load with a fixed phase offset of either 90$^\circ$ or 180$^\circ$ between them. It has the necessary bandwidth, low loss, and amplitude balance needed at the output to ensure adequate signal isolation. The low-noise amplifiers use InP HEMTs in cascaded gain stages. Of all transistors, InP HEMTs have the highest frequency response, lowest noise, and lowest power dissipation. The amplifiers at 30 and 44 GHz use discrete InP HEMTs incorporated into a microwave integrated circuit (MIC). At these frequences, cryogenic MIC amplifiers have demonstrated noise figures of about 10K, with 20\% bandwidth. At 70 GHz, MMICs (Monolithic Microwave Integrated Circuits) architectures, which incorporate all circuit elements and the HEMT transistors on a single InP chip, are used. The LFI will fully exploit both MIC and MMIC technologies at their best. 
For all frequencies, 30--40 dB of gain are sufficient to guarantee that the overall noise is dominated by the front end amplifiers. If additional gain were located in the front end, the power dissipated would grow significantly, putting too much load on the cooler.
Following amplification the signals are passed through a phase switch. The switch consumes microwatts of power, it is broad band, and it works at cryogenic temperatures with switch rates in excess of 1 kHz. The phase switch adds 90$^\circ$ or 180$^\circ$ of phase lag to the signals, thus selecting the input source as either the sky or the reference load at the radiometer output. The phase lagged pair of signals is then passed into a second hybrid coupler, separating the signals. The signals are then transitioned to high performance two meter long bent twisted composite (copper -- stainless steel -- gold-plated stainless steel) rectangular waveguides that carry the double chain signal to the BEU. The waveguides are thermically connected to the three V-grooved shields of the payload at 50, 90, and 140K from the FEU to the BEU, respectively. 

\section{The Back End Unit}
\label{beu}
\subsection{Back end modules}
Each back end module (BEM) comprises two parallel chains of amplification, filtering, detection, and integration. The detected signals are amplified and a low-pass filter reduces the variance of the random signal, providing in each channel a DC output voltage related to the average value. Post-detection amplifiers are integrated into the BEMs to avoid data transmission problems between the radiometer and the electronics box. The sky and reference signals are at different levels, which are equalized after detection and integration by modulating the gain synchronously with the phase switch. Because of the phase switching in the front end modules, a given detector alternately sees the sky and the reference signals. Differences between the detectors are therefore common-mode in the output signals and have no effect on the final difference, which is calculated in the signal processing unit. Each back end module will be packaged, including analog-to-digital converters, into a box of a few centimeters on a side, including the biasing circuitry and the input and output connectors.
\subsection{Data acquisition electronics}
The data acquisition electronics (DAE) comprises the analog conditioning electronics, the multiplexers, the analog-to-digital converters, the parallel-to-serial converters, the control electronics, the communication interface, and the power conditioning and distribution electronics. It performs the following functions: communication with the data processing unit (DPU), including command reception and status transmission; acquisition, conditioning, and multiplexing of the signals; control of the data acquisition chain; transmission of raw data to the signal processing unit (SPU); power supply conditioning and distribution; DC biasing of the FEU and BEU amplifiers; synchronous control of the FEU phase switches; and ON/OFF control of FEU and BEU amplifiers.

\section{The REBA}
\label{reba}
The radiometer electronics box assembly (REBA) comprises three sub-units: the SPU, the DPU, and the power supply unit (PSU).
The REBA supplies all the telemetry and telecommand communication interfaces with the spacecraft, controls the radiometer array assembly through its interface to the BEU, and processes all the radiometer outputs which have been analogue to digital converted in the BEU into science telemetry.

\section{The H$_2$ Sorption Cooler}
\label{scs}
The LFI FEU is cooled to 20K by the hydrogen sorption cooler developed at the JPL (NASA). The operating cooler also provides 18K precooling to the HFI 4K cooler. Each cooler is a Joule-Thomson cooler in which ~0.0065 g/s of hydrogen expands from 5 MPa to ~0.03 MPa through a Joule-Thomson (J-T) expander. The high and low gas pressures are maintained by the fact that the equilibrium pressure of gas above the sorbent bed is a strong function of temperature.

\section{Removal of systematic effects}
\label{syst}

The extreme accurate control of all instrumental systematic effects 
that could in principle affect the measurements (and in practice do
it) is crucial at {\sc Planck} sensitivity levels. On the other hand, 
just the high {\sc Planck} sensitivity allows to successfully apply
several kinds of algorithms to remove systematic effects.
$1/f$ noise and thermal drifts, previously reduced through an accurate
design and realisation of 4~K reference loads and sorption coolers,
will be further subtracted during the data analysis through dedicated
destriping (see, e.g., 
Keihanen et al. 2003
and references therein) 
and map making (see, e.g., 
Natoli et al. 2001
and references therein)
codes respectively in the
TOD domain and on the sky map domain. The first method is blind by 
contruction while the second one requires an accurate noise
parametrization available both from ground testing and from the
in-flight analysis of the TOD. The correction of the effect on the 
{\sc Planck} data from optical distortions, previousloy reduced
through an appropriate optical design study, requires the use of
specific algorithms. External planet transits on the {\sc Planck} 
field of view represents the best way for an accurate main beam 
reconstruction in flight 
(Burigana et al. 2001)
and the main beam distortion effect
on the CMB angular power spectrum recovery can be blindly subtracted
through dedicated deconvolution algorithms jointed to Monte Carlo
simulations for the deconvolved noise subtraction 
(Burigana \& S\'aez 2003)
The straylight effect 
due to the beam far sidelobes is non negligible for {\sc Planck}.
Dedicated ground measures will represent the first step of the
software reduction  of the straylight contamination. 
The study of a semi-blind approach to a joint far beam 
estimate and straylight subtraction is on-going.

\section{The LFI DPC}
\label{dpc}
From the mission operation center (MOC, that will control the {\sc Planck} spacecraft), the scientific data produced by {\sc Planck} will be piped daily to two data processing centres (DPCs). These DPCs will be responsible for all levels of processing of the {\sc Planck} data, from raw telemetry to deliverable scientific products. Although contributions to the LFI DPC, in terms of information on instrument characteristics and prototype software, come from a variety of geographically distributed sites, the operations are mainly centralized in Trieste, where they are run jointly by the OAT and SISSA.
The structure of the LFI (and HFI) DPC has been divided into five levels: LS (during the pre-launch phase, simulation of data acquired from the {\sc Planck} mission on the basis of a software system agreed upon across Consortia), L1 (telemetry processing and instrument control), L2 (data reduction and calibration), L3 (component se\-paration and optimization), and L4 (ge\-neration of final products).

\begin{acknowledgements}
LFI is funded by the national space agencies of the Institutes of the {\sc Planck} Consortium. 
In particular the Italian participation is funded by ASI.
\end{acknowledgements}

\bibliographystyle{aa}

\end{document}